**Мриглод Олеся Ігорівна,**
старший науковий співробітник Лабораторії статистичної фізики складних систем, Інституту фізики конденсованих систем НАН України, кандидат технічних наук
ORCID ID: orcid.org/0000-0003-4415-7061

**Мриглод Ігор Миронович,**
директор Інституту фізики конденсованих систем НАН України, академік НАН України
ORCID ID: orcid.org/0000-0002-0154-9076

# КОЛЕКТИВНЕ АВТОРСТВО В УКРАЇНСЬКІЙ НАУЦІ: МАРГІНАЛЬНИЙ ЕФЕКТ ЧИ НОВЕ ЯВИЩЕ?

Однією з ознак сучасної науки стало формування великих стійких колаборацій науковців, які працюють в рамках проектів, що вимагають концентрації значних матеріальних і людських ресурсів. Результати їхніх досліджень публікуються за авторства колективів, що можуть включати до кількох тисяч імен. Метою нашої роботи є дослідити вплив присутності таких статей на наукометричні індикатори публікаційної діяльності окремих вчених, дослідницьких центрів та усього сегменту української науки. Проведено аналіз бібліометричних даних, що отримані з наукометричної бази Scopus. Показано, що, попри незначну відносну частку публікацій за авторством великих колективів, їхній вплив на наукометричні показники може бути статистично значимим. Отримані результати свідчать про необхідність внесення змін в усталені наукометричні методики та підходи для адекватного врахування цього явища.

*Ключові слова:* колективний автор, наукометрія, групова наука, Україна

**Вступ і мотивація дослідження**

Співпраця — один із наріжних каменів в організації сучасної науки. Взаємодія науковців через обмін інформацією на різних етапах виконання дослідження сприяє вибору актуальної тематики та ефективної методики, дозволяє вдало поєднувати переваги наявного досвіду, а також важлива на стадії аналізу та інтерпретації отриманих результатів. Існують різні форми співпраці, але важко сьогодні уявити науковця, котрий не перебуває у тій чи іншій формі взаємодії зі своїми колегами. Навіть якщо вчений працює формально сам, то на нього неодмінно впливає оточення, на чому, наприклад, наголошується у концепціях так званого т. зв. «мисленнєвого колективу» [1]. Дерек Солла де Прайс – один із творців сучасної наукометрії – використовує термін «невидимий коледж» («invisible college») для окреслення співпраці групи вчених, що виступає однією із домінантних форм організації науки [2]. Подібно до складного пазлу, система знань базується на внеску окремих вчених, кожен із яких опирається на вклад інших. І тут зрозуміло, що, з одного боку, ми «стоїмо на плечах у гігантів», зокрема в стосунках «вчитель-учень», а, з іншого, важливою є також паритетність у відносинах, коли через обмін досвідів та ідей виникає нове знання чи нова ідея. Ці та інші категорії у контексті наукової співпраці є інтуїтивними і давно обговорюються у літературі (див., наприклад, [3]). Збільшенню рівня колаборації сприяє також тенденція сучасної науки до міждисциплінарності, що передбачає залучення до дослідження проблеми різнопланових фахівців [4]. У наукознавстві як відгук на подібні процеси виник новий напрям досліджень «Наука про групову науку» (science of team science – SciTS) (див., наприклад, [5]).

У явній формі співпраця дослідників формалізується у колективному співавторстві. Зрозуміло, однак, що співавторство не цілком адекватно відображає реальний рівень співпраці, бо ще досі у деяких науках (або ж деяких спільнотах) не прийнято, коли співавторами виступають науковий керівник та учень. Так само далеко не завжди, для прикладу, співавторами наукової публікації вказуються інженерно-технічні фахівці, без допомоги яких той чи інший експеримент не вдалося б провести. Окремо можна було б говорити і про порядок авторів у списку співавторів, але мета цього дослідження стосується лише окремого аспекту феномену колективного співавторства, а саме того, що виник порівняно нещодавно як результат появи супервеликих міжнародних проектів[1]. Проте, перед тим як чіткіше окреслити предмет нашого дослідження, проаналізуємо картину співавторства, яка ще донедавна вважалася типовою для більшості наук.

На Рис.1 показані нормовані частотні розподіли публікацій за числом співавторів як в окремих наукових періодичних виданнях, так і в певних їх групах, що об'єднують журнали окремих Відділень Національної академії наук (НАН) України. Зокрема, такі розподіли побудовані для: «Журналу фізичних досліджень» (дані про 962 публікації за 1996–2016 роки) [6], журналу «Condensed Matter Physics» (1282 публікації за 1993–2017) [7], «Мінералогічного журналу» [8] та даних про видання трьох Відділень НАН України, а саме економіки (1787 публікацій, 2012–2016), хімії (1563, 2013–2016) та наук про Землю (1796, 2011–2016). Як бачимо на Рис. 1, отримані залежності доволі схожі, хоча виділяються серед них криві, що відповідають Відділенню економіки та Відділенню хімії. Максимум на частотній кривій для видань Відділення економіки спостерігається для одноосібних публікацій. Для усіх інших випадків максимум зміщується в сторону більшого числа співавторів, що особливо чітко видно для видань Відділення хімії НАН України. Таким чином, бачимо, що найбільш імовірні розміри колективів співавторів дещо різняться для різних наук (наборів даних), зокрема: одноосібні статті виразно лідирують в економічних виданнях, фізичні публікації найчастіше написані двома авторами, а хімічні – у групі 3-4 співавторів. Звичайно, такі значення є актуальними лише із прив'язкою до конкретного набору даних і певного часового періоду, але вони цілком відображають загальну тенденцію в науці до колективного співавторства, що почала спостерігатися у XX ст. [4], [5].

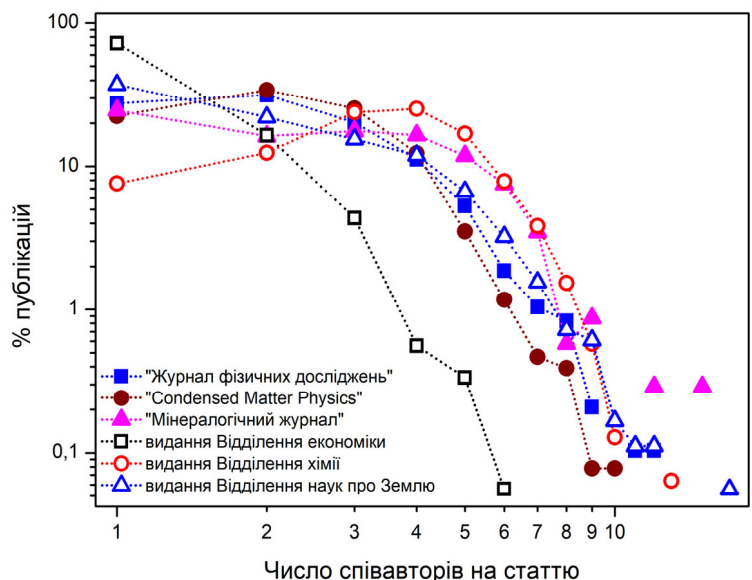

*Рис. 1. Нормовані частотні розподіли публікацій за числом співавторів на роботу для:*

---

[1] Тут доречно пригадати в якості чи не першого прикладу стійкого колективного співавторства Ніколя Бурбакі з містом афіліації Нанкаго (Нансі + Чикаго). Цей колективний псевдонім був запроваджений у 30-х роках минулого століття групою математиків, які особливо активно працювали в 50-60-х роках і мали на меті видання серії книг, що відображали б сучасний на той час стан математики.

«Журналу фізичних досліджень» [6], журналу «Condensed Matter Physics» [7], «Мінералогічного журналу» [8] та сукупні дані про видання трьох Відділень НАН України (економіки, хімії та наук про Землю).

Такий процес є цілком природнім і другою типовою його ознакою, окрім зсуву положень максимумів, можна вважати типову форму хвостів відповідних розподілів (зокрема тих, що приведені на Рис. 1): у всіх випадках спадання близьке до пуасонівського.

Інший тип колективної наукової співпраці виникає при реалізації масштабних і тривалих у часі проектів, що мають на меті отримання та обробку великих масивів інформації, як того часто вимагають сучасні експерименти. Для реалізації таких проектів вимагається концентрація значних коштів, поєднання зусиль великих колективів дослідників і технічних працівників, які представляють іноді десятки чи, навіть, сотні країн. Виникає так звана «велика наука» («Big Science»), що відкриває неймовірні можливості для унікальних у своєму роді досліджень, проте платою стає складність надійної перевірки результатів шляхом їх відтворення в інших умовах [2], [5]. І як наслідок, виникає також потреба у переосмисленні таких базових понять у науці як авторство та, зокрема, у внесенні змін у вже усталені методології його врахування при спробах формалізації та оцінювання наукового продукту.

Пригадаємо, для прикладу, міжнародний проект "Геном людини" Помилка! Джерело посилання не знайдено., мета якого полягала в ідентифікації усіх пар основ, що складають людську ДНК, та визначенні всіх генів людини з фізичної та функціональної точок зору, що офіційно розпочався в 1990 р. і розраховувався на 15 років. Фактично проект було завершено 14 квітня 2003 р., коли анонсували, що відомий майже повний геном. Інформація щодо послідовності останньої хромосоми була опублікована у травні 2006 р. Цей проект досі залишається найбільшим у світі біологічним проектом (витрати за проектом склали близько 5 млрд дол.), що фінансувався урядом США через Національний інститут охорони здоров'я (NIH), а також численними іншими групами з усього світу. Більшість досліджень виконувалася у двадцяти університетах та науково-дослідних центрах США, Великобританії, Японії, Франції, Німеччина та Китаю.

Інший приклад масштабного міжнародного проекту стосується уже фізики. Мова йде про Великий адронний колайдер [10][9] (див. також [11]) — найбільший у світі прискорювач заряджених частинок на зустрічних пучках, що був створений у Європейському центрі ядерних досліджень (CERN), поблизу Женеви (Швейцарія). До його будівництва (1998-2008 рр.) та встановлення відповідного обладнання було залучено понад 10 тис. фахівців, які представляли сотні дослідницьких центрів із понад 100 країн світу. Це найбільша та найдорожча (бюджет оцінюється у 9 млрд. дол) в світі машина і вимірювальна установка водночас, яка розміщена в тунелі кільцевої форми довжиною 27 км на глибині від 50 до 175 метрів. Основним мотиватором для розбудови таких високовартісних інструментів є пошук відповідей на питання, що стосуються нової фізики. Беруть участь у цих дослідженнях й українські науковці, зокрема з ННЦ "Харківський фізико-технічний інститут", Інституту сцинтиляційних матеріалів НАН України, Інституту теоретичної фізики ім. М.М.Боголюбова НАН України та ін.

На відміну від згаданих вище форм "природньої" колективної співпраці (робота в одній чи кількох формальних або ж неформальних групах, стосунки «вчитель-учень» тощо), у випадку міжнародних суперпроектів мова вже йде про формування стійких, певним чином обумовлених, колаборацій науковців (колективів співавторів у випадку публікації отриманих результатів) задля реалізації тривалої програми досліджень. Якщо в утворенні та існуванні перших домінує соціальна або комунікаційна складова (див, напр., [12], [13]), то другі створюються скоріше організаційно та діють формально, що не передбачає необхідності включення соціальних взаємодій [14]. Відповідно, остання обставина розглядається іноді як підстава для сумніву щодо коректності трактування співавторства у рамках великомасштабних проектів як відображення власне наукової співпраці [15].

Очевидно, що публікаційна діяльність великих колаборацій спричиняє необхідність внесення суттєвих змін у вже усталені методики, що стали традиційними в наукометрії, яка, опираючись на бібліографічну інформацію, статистику цитувань та інші джерела, оперує як правило поняттями «типових», «середніх» або «нормальних» показників. І хоча зараз добре відомо, що розподіли в сфері наукової діяльності зазвичай відмінні від нормального, певні усереднені з їх допомогою значення можуть використовуватися для різних задач у якості відправної точки для порівняння. Так, із врахуванням дисциплінарних або ж інших особливостей конкретної вибірки даних, можна говорити про «нормальний» рівень самоцитувань або ж прийнятну частку нецитованих статей, про характерний розмір авторських колективів тощо. Однак, ситуація кардинально змінюється (а в ряді випадків втрачається сенс на рівні означень), якщо до сукупності даних, що містять результати публікаційної діяльності лише колективів з "природньою" співпрацею, додати ще й дані великих колаборацій. Для прикладу, проведені нещодавно дослідження впливу процесів наукової колаборації свідчать про якісні відмінності у динаміці формування таких дослідницьких груп і їх еволюції, зокрема показано, що розміри колективів авторів у масштабних проектах розподілені за степеневим законом [14], [16]. Маємо також врахувати, що з початком XXI ст. сформувалося досить багато стійких у певних часових рамках колаборацій, які активно публікуються і можуть привносити статистично значущий вклад у показники загальної публікаційної діяльності. Зауважимо також, що попри доволі вузьку наукову спрямованість супервеликих міжнародних проектів, великі колаборації співавторів можуть давати також статистично значущий вклад у загальну публікаційну картину країн, наукових товариств чи великих наукових центрів, де виконуються дослідження за широким спектром наук. Тому важливими і актуальними стають питання про внесення відповідних змін у традиційні наукометричні методики, бо за таких обставин вклад цієї групи робіт не може розглядатися як нехтовно малий або ж флуктуаційний. Очевидно, що найпростіший шлях — це виділення їх в окрему категорію із наступною модифікацією відповідних методик. Підставою для цього є й те, що такого роду співавторство не лише відрізняється від звичного чи "природного" (зокрема, через невиконання загальноприйнятих критеріїв співавторства, наприклад, так звані Ванкуверські критерії [17]), але й формує доволі специфічну динаміку публікаційної активності та акумуляції цитувань. Тому великі наукові колаборації слід розглядати окремо і вони вже сьогодні трактуються як особливі «гравці» на наукометричному полі одним із найавторитетніших наукометричних сервісів Web of Science (WoS). Так, WoS у своєму пошуковику виділяє окреме поле «Group Author», що реалізує пошук стійких колаборацій колективних співавторів із своєю специфічною назвою. Відповідно, за цією назвою відтворюється повний перелік прізвищ індивідуальних авторів, якщо останній є великим (більше 400) [18].

Головна мета цієї роботи — це аналіз присутності та ролі колективного авторства в публікаційній діяльності українських науковців, визначенні його впливу на позиціонування вітчизняної науки на світовій арені, а також на наукометричні профілі наукових установ в Україні та наукометричні показники окремих вчених. Для цього будуть використані бібліометричні дані з бази Scopus.

**Коротка інформація про використані дані**
Для проведення дослідження, в залежності від цілей та поставлених завдань, нами використовувалися декілька наборів даних, що були отримані з бази даних Scopus станом на початок 2020 року, а саме результати пошуку документів:
  i.  за ключовим словом «Ukraine» у полі «AFFILCOUNTRY» (це 349 різних афіліацій);

ii. за профілями 72 установ НАН України – усіх з переліку [19], які на момент пошуку такі профілі мали;
iii. за профілями вибраних українських вчених (всього більше 50).

Зведена інформація про дані (і) приведена у Табл. 1.

*Табл. 1. Характеристики набору даних (і) про публікації, в афіліаціях яких фігурували назви українських наукових установ. Дані отримані з бази даних Scopus станом на початок 2020р.*

| Часовий період | 1889–2020 | 1889–1990 | 1991–2020* |
|---|---|---|---|
| Загальна кількість публікацій | 243 581 | 35 208 | 208 373 |
| Середня кількість авторів на одну роботу | 15.5 | 2.8 | 17.6 |
| Медіана кількості авторів на одну роботу | 3 | 3 | 3 |
| Частка одноосібних робіт | 17% | 22% | 16% |
| Максимальна кількість авторів на одну роботу | 5214 | 64 | 5214 |
| Середня кількість цитувань на одну роботу | 6.2 | 2.7 | 6.8 |
| Медіана цитувань на одну роботу | 1 | 0 | 1 |
| Максимальна кількість цитувань на одну роботу | 5359 | 1668 | 5359 |
| Частка нецитованих робіт | 43% | 64% | 39% |

*при підрахунку показників цитованості 2019 та 2020 роки не враховувалися з огляду на інертність акумуляції цитувань.*

Загальна кількість публікацій у базі даних Scopus, що віднесені сервісом до України, стрімко зростає, починаючи з кінця 1950-х років (див. Рис. 2) — від десятків статей в рік і до десятка тисяч (більше 13.5 тис.) у 2019 р. З врахуванням мети нашого дослідження, що стосується, насамперед, феномену появи стійких колаборацій авторів, які почали формуватися наприкінці XX ст., ми розділили (подібно як у [16]) дані за двома часовими періодами — до 1990 року включно та від 1991 до 2020. Як бачимо з Табл. 1, і це природно з огляду на офіційне становлення незалежної України лише у 1991, переважна більшість публікацій, що асоціюються у базі Scopus з Україною, опубліковані після 1990 р. (понад 85.5%). 113 публікацій, що складає менше 0.05%, знаходимо до 1950 року.

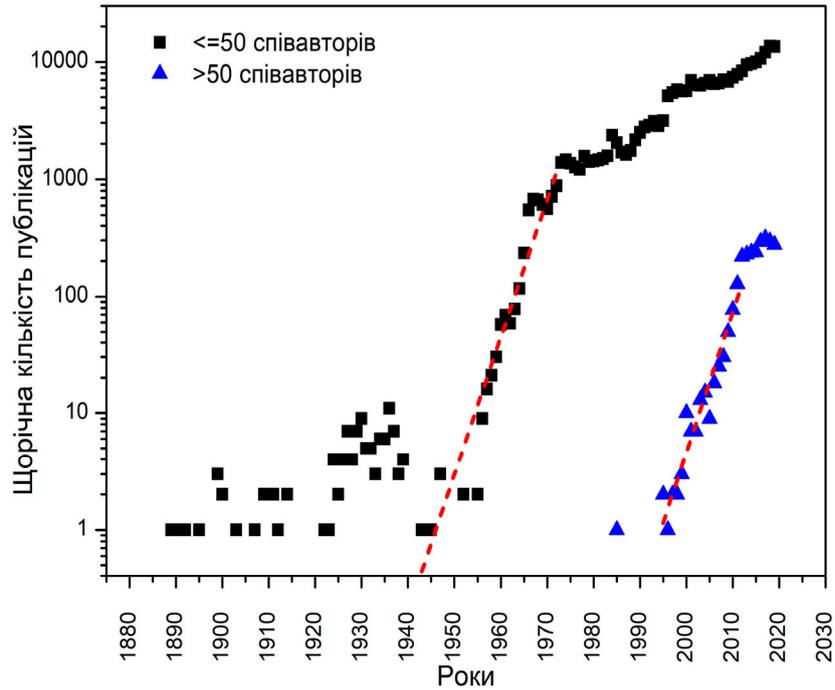

***Рис. 2****. Щорічні дані про опубліковані статті, де в полі афіліацій згадується Україна (або ж установи на території сучасної України): а) тих, у яких розміри авторських колективів не перевищували 50 осіб (квадратики); б) усі інші (трикутники).*

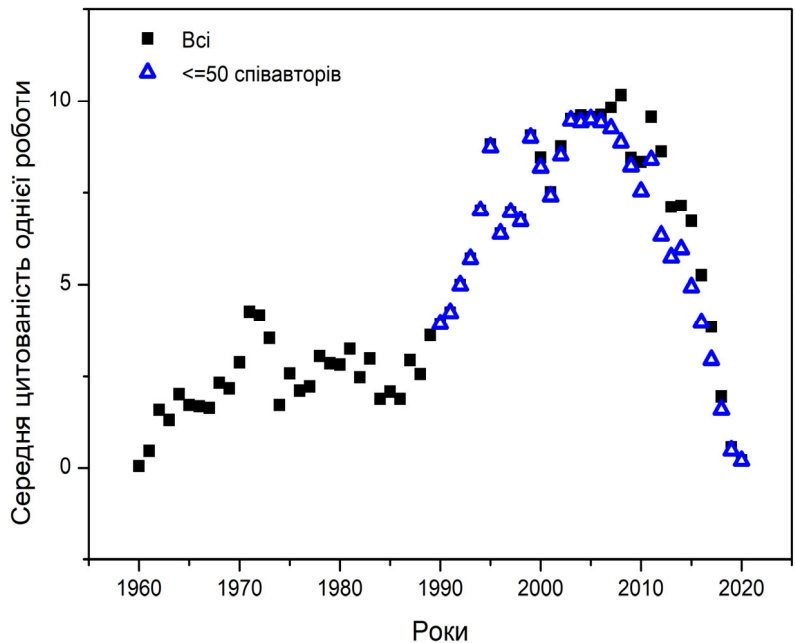

***Рис. 3.*** *Щорічні значення середнього числа цитувань, отриманих на одну статтю, що слідує з набору даних (і): квадратиками показано для випадку, коли враховано всі роботи, а трикутниками – коли враховано лише роботи, де кількість співавторів не перевищує 50 (з 1990 року).*

Зауважимо також, що перші публікації, які віднесені у базі Scopus до України, фіксуються з 1889 р. У найраніших серед них як місце праці автора згадується нинішній Одеський національний університет імені І.І.Мечникова. Цікаво, що переклад тексту лекції Іллі Мечникова, яку він прочитав у грудні 1890 р. в Інституті Пастера, також увійшов до цього переліку:

- Heritsch, A., 1889. Ueber das allgemeine Gesetz der bei dem Lösen von Salzen im Wasser auftretenden Volumenverminderung. Annalen der Physik, 272(1), pp.115-122. *[У афіліації автора вказано «Physik. Lab. der Univ. Odessa», що відноситься до сучасного Одеського національного університету імені І. І. Мечникова]*
- De Metz, G., 1890. Ueber die compressibilität der Oele und Colloide1. Annalen der Physik, 277(12), p. 663-674. *[У афіліації фігурує «Phys. labor. d. Univ. Odessa», що відноситься до сучасного Одеського національного університету імені І. І. Мечникова]*
- Metschnikoff, E., 1891. Lecture on phagocytosis and immunity. British medical journal, 1(1570), p.213. *[у афіліації: «Chef de Service de l'Institut Pasteur, Paris; late Professor of Zoology in the University of Odessa]*

Кілька слів про найцитованіші публікації. Як бачимо, з Табл. 1 найцитованіша робота станом на січень 2020 р. набрала 5359 цитувань (уже 5475 на кінець квітня 2020 р.). Ідеться про статтю «Observation of a new boson at a mass of 125 GeV with the CMS experiment at the LHC», опубліковану у 2012 р., у журналі «Physics Letters B» групою з 2891 співавторів, які працювали в рамках CMS колаборації. Наступними з топ п'яти найцитованіших робіт є такі:

- Granger, C.B., Alexander, J.H., McMurray, J.J., Lopes, R.D., Hylek, E.M., Hanna, M., Al-Khalidi, H.R., Ansell, J., Atar, D., Avezum, A. and Bahit, M.C. et al., 2011. Apixaban versus warfarin in patients with atrial fibrillation. New England Journal of Medicine, 365(11), pp.981-992. *[Колаборація ARISTOTLE Comm Investigators, 32 співавтори, 4866 цитувань у січні 2020]*
- Abubakar, I.I., Tillmann, T. and Banerjee, A. et al., 2015. Global, regional, and national age-sex specific all-cause and cause-specific mortality for 240 causes of death, 1990-2013: a systematic analysis for the Global Burden of Disease Study 2013. Lancet, 385(9963), pp.117-171. *[в рамках проекту GBD, 717 співавторів, 3523 цитувань ]*
- Collaboration, C.M.S., 2008. The CMS experiment at the CERN LHC. Journal of Instrumentation, 3(8), S08004 *[колаборація CMS, 3084 співавтори, 3503 цитування]*
- Kovtyukhova, N.I., Ollivier, P.J., Martin, B.R., Mallouk, T.E., Chizhik, S.A., Buzaneva, E.V. and Gorchinskiy, A.D., 1999. Layer-by-layer assembly of ultrathin composite films from micron-sized graphite oxide sheets and polycations. Chemistry of materials, 11(3), pp.771-778. *[7 співавторів, 2632 цитування]*

Таким чином, 4 з 5 найцитованіших публікацій України є продуктом діяльності великих колаборацій. Водночас, для найцитованіших робіт, опублікованих до 1990 року, характерною є невелика кількість співавторів:

- Kukhtarev, N.V., Markov, V.B., Odulov, S.G., Soskin, M.S. and Vinetskii, V.L., 1995. Holographic storage in electrooptic crystals.: I. steady state. In: Landmark Papers On Photorefractive Nonlinear Optics (pp. 37-48). *[5 співавторів, 1668 цитувань]*
- Brack, M., Damgaard, J., Jensen, A.S., Pauli, H.C., Strutinsky, V.M. and Wong, C.Y., 1972. Funny hills: The shell-correction approach to nuclear shell effects and its applications to the fission process. Reviews of Modern Physics, 44(2), p.320. *[6 співавторів, 1336 цитувань]*
- Kivshar, Y.S. and Malomed, B.A., 1989. Dynamics of solitons in nearly integrable systems. Reviews of Modern Physics, 61(4), p.763. *[2 співавтори, 1212 цитувань]*
- Volkov, D.V. and Akulov, V.P., 1973. Is the neutrino a Goldstone particle? Physics Letters B, 46(1), pp. 109-110. *[2 співавтори, 1054 цитування]*
- Ivakhnenko, A.G., 1971. Polynomial theory of complex systems. IEEE transactions on Systems, Man, and Cybernetics, (4), pp.364-378. *[1 автор, 912 цитувань]*

Загалом у списку 38 найцитованіших статей, кожна з яких отримала понад 1000 цитувань, знаходимо лише 4 опубліковані до 1990 р., а 24 — після 2005 р. Половина цих робіт, а саме 16, має авторський колектив, в якому понад 50 співавторів. У ще 9 статтях колектив співавторів складає від 10 до 50 осіб, а в 13 публікацій він є меншим за 10 (серед них дві одноособові публікації та 4 статті, в яких по два співавтори).

На Рис. 3 приведені дані про середню цитованість на одну статтю з прив'язкою до року її публікації. Виразно бачимо ріст середнього числа цитувань, що починається перед 1990 р. До певної міри такий ріст подібний (і описується практично такою ж залежністю) до того, що спостерігається на Рис. 2 для росту числа публікацій у 1950-х роках (крива з чорними квадратиками). Кореляцію між цими процесами можна було б досліджувати окремо, а в контексті завдань цієї роботи важливо звернути увагу на свого роду максимум на Рис. 3, що спостерігається в інтервалі 2000-2012 роки. З одного боку, він є досить типовим, оскільки процес цитувань доволі інертний і тому роботи останніх років в середньому будуть завжди менш цитованими. З іншого боку, як побачимо далі, публікації стійких колаборацій мають суттєво вищі показники цитувань, а отже потенційно можуть впливати на загальну усереднену картину. До обговорення цього питання ми ще повернемося згодом.

Про набори даних (ii) та (iii) більш детально будемо говорити далі.

**Аналіз даних: публікації колаборативних авторів в Україні**
Одне із найважливіших наших завдань полягає у тому, щоб знайти спосіб виділити з набору даних (i) публікації, що належать авторству великих колаборацій. Як один із можливих підходів до вирішення цієї проблеми, скористаємося на першому етапі з аналізу відносної кількості публікацій за числом співаторів у цих публікаціях.

На Рис. 4 показано частотний розподіл публікацій за числом авторів для двох публікаційних періодів, а саме до 1990 року включно та за 1991–2020 рр. У подвійному логарифмічному масштабі видно, що крива для першого періоду має форму, схожу до кривих, які були приведені на Рис. 1, із характерним хвостом, що типовий для "природніх" колективів співавторів, на фоні якого спостерігаються певні флуктуаційні значення. Очевидно, що переважна більшість публікацій знаходиться в області малих значень і характеризуються вони невеликими розмірами авторських груп (від 1 до 4), а далі спостерігаємо різкий спад. Натомість для другого періоду часу (кружечки на Рис. 4) спостерігаємо процес формування довгого хвоста, тобто розподіл починає набувати рис типових для безмасштабного розподілу. Однак, оскільки статистика все ж є ще недостатньою, зокрема й тому, що феномен стійких колаборацій є порівняно недавнім, а окрім того, його прояви слід швидше очікувати лише у певних науках, то висновки про якісь загальні риси такого типу розподілу на основі даних (i) робити зарано. Водночас, можна чітко бачити, що у другому випадку в області великих розмірів авторських колективів спостерігаються доволі різкі окремі максимуми, які не можуть інтерпретуватися як флуктуаційні.

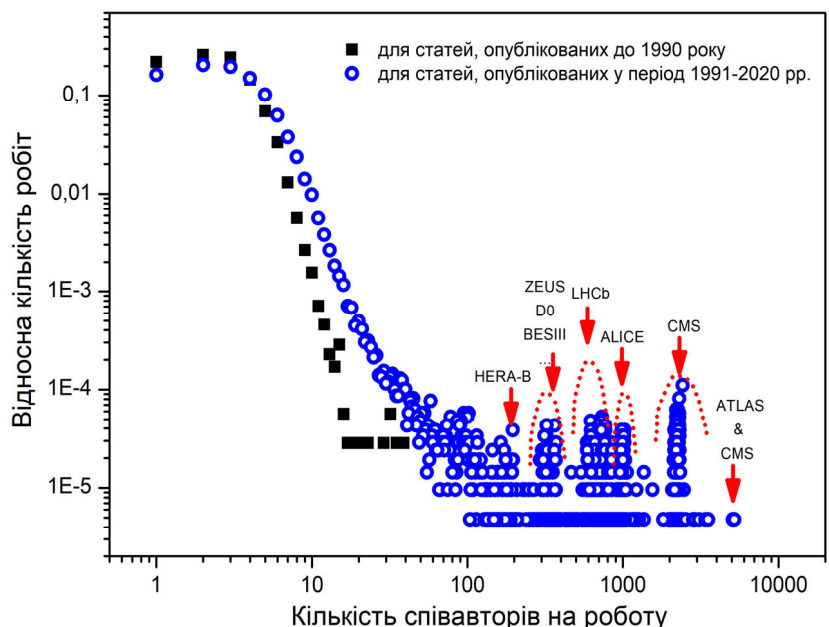

***Рис. 4.*** *Нормований частотний розподіл публікацій українських авторів за числом співавторів. Наведено результати для двох публікаційних періодів: 1889–1990 (квадратики) та 1991–2020 (кружечки). Абревіатури – назви колаборацій.*

У Табл. 2 приведені дані щодо кількості публікацій (абсолютні і питомі значення) із розмірами авторських колективів, що перевищують певну задану величину $n$, починаючи з $n = 50$, для даних (і) із другого часового періоду.

***Табл. 2.*** *Абсолютні значення числа публікацій із розмірами авторських колективів, що перевищують певні задані значення, та їхні питомі частки у загальній кількості публікацій для двох вибірок із набору даних (і). Також приведено дані щодо їх цитувань.*

| Фіксована кількість співавторів $n$ | Число публікацій, у яких кількість співаторів перевищує $n$ | Частка (в %) публікацій, у яких кількість співаторів перевищує $n$ | Число цитувань публікацій, у яких кількість співаторів перевищує $n$ | Частка (в %) цитувань публікацій, у яких кількість співаторів перевищує $n$ |
|---|---|---|---|---|
| 50 | 2511 | 1.03 | 143027 | 9.5 |
| 300 | 1920 | 0.79 | 114484 | 7.6 |
| 500 | 1755 | 0.72 | 108730 | 7.2 |
| 1000 | 1067 | 0.44 | 58313 | 3.9 |
| 2000 | 927 | 0.38 | 50767 | 3.4 |

Зауважимо, що для періоду до 1990 р. (як видно з Рис. 4) практично усі колективи співавторів є розміру, меншого за 50 осіб. Для другого часового періоду знаходимо, що частки публікацій з $n>50$ та $n>2000$ рівні 1.03% та 0.38%, відповідно. Здавалося б це досить малі величини, якими можна було б знехтувати, проте можемо легко переконатися в хибності такого підходу. Перший сигнал на статистичну значущість частки статей із надвеликими колективами співавторів дає порівняння середнього числа співавторів для двох часових періодів з Табл. 1, де знаходимо значення 2.8 та 17.6, відповідно. Другий приклад, який слід тут навести, стосується уже іншої важливої в наукометрії величини — мова йде про рівень цитувань наукових публікацій. Як бачимо з Табл. 2, статті, що опубліковані колективами співавторів з розміром понад 2000 осіб і складають всього лише 0.38% від загального числа публікацій з другого часового періоду, забезпечують 3.4% усіх їхніх цитувань. Тобто ігнорувати ці публікації з огляду на їх статистичну значущість абсолютно неприпустимо. Щоб посилити це твердження, наведемо ще одну цікаву цифру, яка стосується рівня цитованості. Так, із переліку найвпливовіших публікацій із рівнем цитувань понад 1000, більше 42% становлять публікації, у яких понад 150 співавторів. Також можемо бачити на Рис. 3, що після 2005 року статті, написані у великих колабораціях (понад 50 співавторів), роблять суттєвий внесок у значення середньої цитованості на одну роботу. Таким чином, уже на цьому етапі дослідження можна стверджувати, що вклад публікацій, які асоціюються із великими колективами співавторів, на показники рейтингування української науки є доволі помітним.

Спробуємо тепер конкретизувати те, про які стійкі колаборації і які надколективи співавторів йдеться, коли ми говоримо про чіткі піки, що спостерігають у хвості розподілу (див. Рис. 4) для *n>500*. В області локалізації піків можемо провести аналіз заголовків публікацій, що часто достатньо для встановлення приналежності публікації до циклу робіт в рамках певного проекту, а додатковий пошук в «ручному режимі» (перегляд анотацій, маркування колективного автора у базі Web of Science, тощо) забезпечує остаточну перевірку. Беручи до уваги типовий розмір авторських груп в рамках того чи іншого проекту (з певними невеликими модифікаціями) та враховуючи часовий проміжок, можна визначити колективних авторів для окремих циклів робіт. Таким чином, було виявлено (див. Рис. 5 далі), що найбільші групи співавторства (понад 5 тис. осіб) фігурують у чотирьох роботах, що виконувались спільно у рамках двох великих експериментів, які пов'язані з великим адронним колайдером (англ. Large Hadron Collider, LHC) Європейського центру ядерних досліджень (CERN) [8], а саме: ATLAS [https://atlas.cern] та CMS [https://cms.cern]. При цьому були задіяні співавтори з двох академічних установ України — ННЦ "Харківський фізико-технічний інститут" та Інституту сцинтиляційних матеріалів. Далі, в порядку зменшення розміру авторського колективу, слідує невелика кількість публікацій астрономічного (низка колективів співавторів, що включають понад 3.5 тис. прізвищ) і медичного (дослідження COMPASS) спрямування, а також велика група публікацій з результатами експерименту в рамках проекту CMS, що, як правило, пов'язані з уже згаданими вище установами ННЦ "Харківський фізико-технічний інститут" спільно з Інститутом сцинтиляційних матеріалів. Така своєрідна «монополія» простягається фактично на понад 900 статей, кожна із яких опублікована групами від приблизно 2 тис. до 3 тис. співавторів. У діапазоні між близько 900 та понад 1100 осіб знаходимо більше 270 публікацій, переважна більшість із яких відноситься до Інституту теоретичної фізики ім. М.М.Боголюбова НАН України, що бере участь у ще одному проекті CERN на LHC під назвою ALICE [http://alice-collaboration.web.cern.ch/]. Далі виявляється група із близько 500 публікацій, що асоціюється з авторським колективом розміром приблизно 540 осіб, що відображають участь ННЦ "Харківський фізико-технічний інститут" та Інституту ядерних досліджень в експерименті CERN під назвою LHCb [https://home.cern/science/experiments/lhcb]. Якщо рухатися у порядку спадання розмірів колективів співавторів, то знаходимо також статті, які відображають роботи групи D0 [https://www-d0.fnal.gov/] в рамках експерименту ZEUS [https://www-zeus.desy.de/], що проводився на базі іншого колайдера – HERA [11]. Перелік задіяних установ при цьому розширюється і знаходимо тут уже вищі навчальні заклади (для прикладу, КНУ імені Тараса Шевченка). Окрім колаборативних робіт із фізики, можна виявити також публікації медичного спрямування. Скажімо, більше трьох десятків робіт, що мають в середньому 650 співавторів, були опубліковані в рамках проекту Global Burden of Disease Study (GBD) у період 2014–2017 рр. Астрофізичні та астрономічні колаборативні публікації переважно характеризуються меншими авторськими групами порядку 100–150 осіб, зокрема: DarkSide [http://darkside.lngs.infn.it/], Borexino [http://borex.lngs.infn.it/] та EURECA [http://www.eureca.kit.edu/].

Не у всіх випадках спроба ідентифікації стійких надколективів співавторів приводить до однозначних результатів. Цікавим тут є приклад цілої низки публікацій, що стосуються експерименту BESIII (Beijing Spectrometer) [http://bes3.ihep.ac.cn/], які сервісом Scopus відносяться до профіля Інституту теоретичної фізики ім. М.М.Боголюбова НАН України, тоді як Web of Science взагалі не пов'язуються ці роботи з Україною (один із співавторів вказує як афіліацію науковий центр Об'єднаний інститут ядерних досліджень у Дубні (Росія) із приміткою «on leave from» з Інституту теоретичної фізики ім. М.М. Боголюбова НАН України). У цьому ж контексті в якості контрприкладу можна навести одну із публікацій великої колаборації (3529 осіб), прив'язану до НДІ «Кримська астрофізична обсерваторія», яка за версію Web of Science додає 854 цитувань у «скарбничку» України, проте сервісом Scopus відноситься до Росії. Таким чином, можна знайти певні невідповідності між даними, якщо співставляти перелік публікацій, що віднесені до України за афіліаціями, та індивідуальні переліки публікацій для кожного з співавторів зокрема.

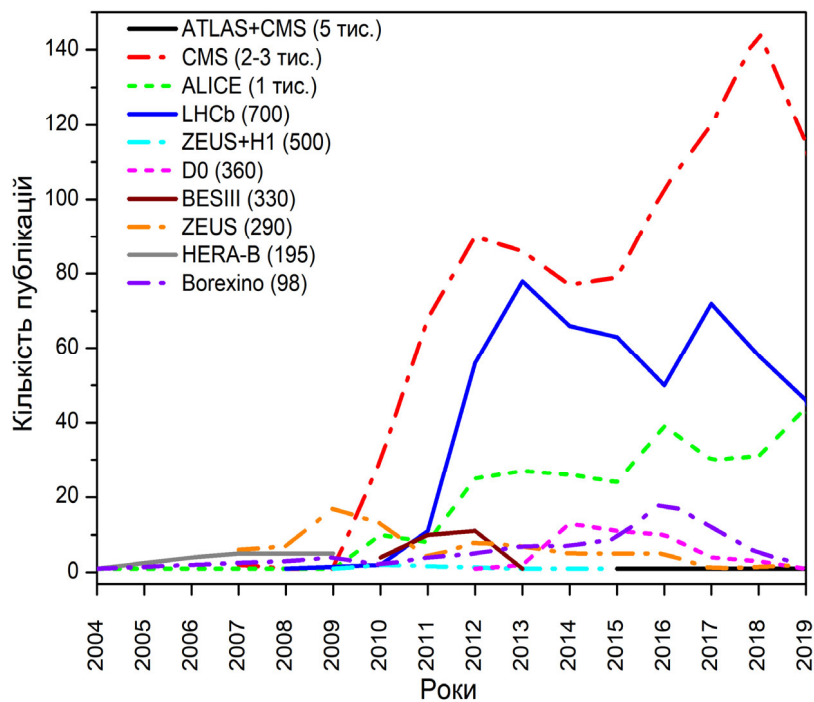

***Рис. 5.*** *Кількість публікацій деяких основних колективних авторів[2] з розбивкою по роках, які за афіліаціями віднесено до України системою Scopus. В дужках вказано середній розмір авторських колективів.*

Певні уявлення про публікаційну динаміку порівняно стійких колективів співавторів, які працювали над реалізацією спільних проєктів, з розбивкою по роках дає Рис. 5. Список основних колективних авторів наведено у порядку спадання розмірів авторських колективів. Для різних колаборацій можна спостерігати певну немонотонну поведінку з піками у різні роки. Очевидно, що вона відображає специфічні часові інтервали, що пов'язані з особливостями виконання того чи іншого проєкту, з яким колективний автор асоціюється. Так, для прикладу, у поведінці трьох кривих, що розміщені найвище на Рис. 5, можна знаходити кореляції як з деякими історичними подіями у співпраці України та Європейського центру ядерних досліджень (підписання спільної декларації, 2011 р.; набуття статусу асоційованого члена

---

[2] Приналежність робіт до колективного автора здійснювалась у ручному режимі, тому отримані дані не претендують на абсолютну повноту і можуть містити певні неточності.

CERN, 2016 р.), так і з планом виконання ремонтних робіт на Великому адронному колайдері в 2013-2015 рр.

Таким чином, бачимо, що участь України у великих наукових проектах, зокрема на Великому адронному колайдері, це важливий фактор, що забезпечує ріст міжнародного авторитету нашої країни та рівня її присутності у міжнародних наукових рейтингах.

**Колаборативні автори та наукометричні профілі установ**

Розглянемо тепер питання про вплив публікацій за авторством великих колаборацій на наукометричний профіль установ. Розпочнемо з даних про 7686 статей, що їх Scopus відносить до профіля *ННЦ "Харківський фізико-технічний інститут" (ХФТІ)*. Розподіл за кількістю співавторів для цієї чи не найбільшої інституції НАН України доволі близько нагадує ту залежність, що показана на Рис. 4. У Табл. 3 приведені деякі з базових показників, які у тих чи інших цілях традиційно розраховують для наукової установи, зокрема: середньорічна кількість публікацій та цитувань, середня кількість цитувань на одну публікацію, число та/або відсоток нецитованих робіт, груповий індекс Гірша [20]. Варто зауважити, що груповий індекс Гірша може використовуватися для аналізу діяльності окремих дослідницьких груп чи установ, проте цей інструментарій потребує модифікації, коли мова йде про порівняння між собою груп чи установ різного розміру, див. [21], [22]. У Табл. 3 також приведено значення усіх вищезгаданих показників, але вже для випадку, коли не враховуються статті великих колективів співавторів.

*Табл. 3. Наукометричні показники для публікацій ННЦ "Харківський фізико-технічний інститут", що відносяться до профілю установи у базі Scopus станом на кінець січня 2020 р. Бралися до уваги роботи, опубліковані в період від 1991 до 2019 року включно.*

| Показник | Усі дані | Для окремих вибірок даних, що не включають статті, де кількість співавторів перевищує *n* | | | |
|---|---|---|---|---|---|
| | | *n*=5000 | *n*=1000 | *n*=500 | *n*=50 |
| Загальна кількість публікацій | 7359 | 7355 | 6435 | 5932 | 5804 |
| Частка неврахованих статей | - | 0.05% | 13% | 19% | 21% |
| Середня кількість співавторів на одну роботу | 338 | 335 | 63 | 8 | 5 |
| Середня кількість публікацій на рік | 254 | 254 | 222 | 205 | 200 |
| Середня кількість цитувань на рік | 3730 | 3687 | 2059 | 1382 | 1143 |
| Середня кількість цитувань на одну роботу | 14.7 | 14.5 | 9.3 | 6.8 | 5.7 |
| Частка нецитованих робіт | 30% | 30% | 34% | 36% | 37% |
| Індекс Гірша | 120 | 119 | 91 | 75 | 62 |

Як бачимо з Табл. 3, порівняно невелика частка статей у великих колабораціях може помітно впливати на загальні наукометричні показники. Для прикладу, якщо вилучити з списку публікацій роботи колективного авторства з розміром групи тисячу і більше осіб, доля яких складає майже 13%, то фактично на 45% зменшується показник середньорічної цитованості статей установи та більше ніж на третину падає середня кількість цитувань на одну публікацію. І навіть такий показник, як індекс Гірша – досить консервативний в ділянці великого числа публікацій – зменшується на 29 одиниць (або майже на 25%) при вилученні цих публікацій (див. Табл. 3). Причиною цьому є висока цитованість (включаючи високу самоцитованість) публікацій, що належать до однієї групи зі спільним колективним автором.

Таким чином, бачимо, що розгляд в рамках одного масиву даних як публікацій із «природнім» співавторством, так і опублікованих в результаті великих стійких колаборацій, може приводити до кардинально відмінних результатів при розрахунку усереднених величин. На це ми вже звертали увагу раніше, і причина тут цілком зрозуміла, бо розрахунок середніх має сенс лише тоді, коли випадкові величини описуються одним розподілом, який близький до нормального. Про те, що це не є так у нашому випадку, свідчить як частотний розподіл публікацій за числом співавторів (див., для прикладу, Рис. 4), так і частотний розподіл публікацій за їх цитованістю, що приведений для профілю ННЦ "Харківський фізико-технічний інститут" на Рис. 6.

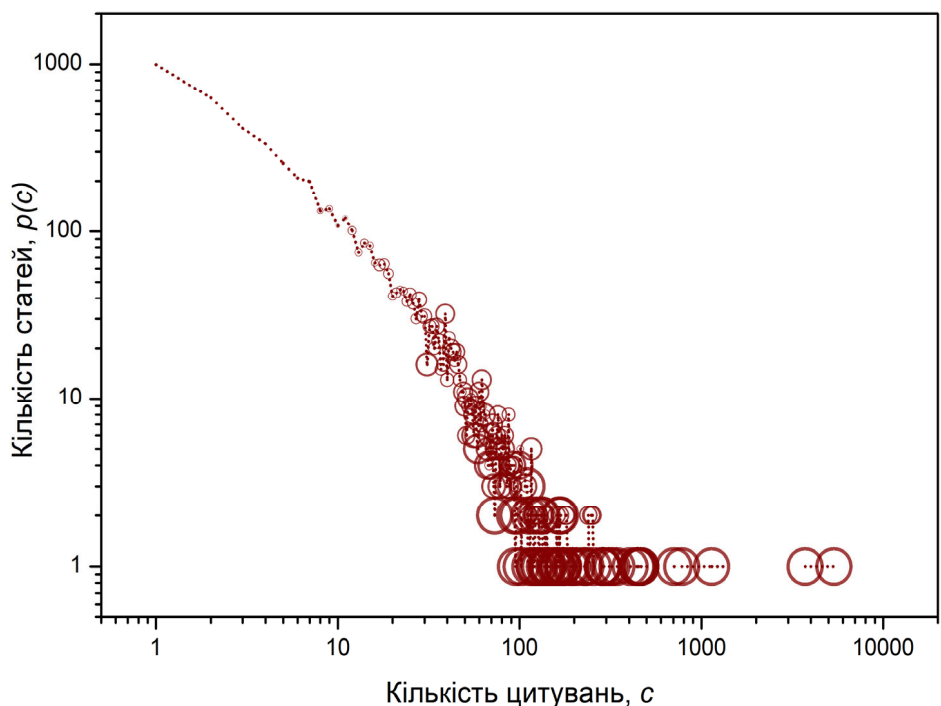

*Рис. 6. Частотний розподіл публікацій для профілю ННЦ "Харківський фізико-технічний інститут" у Scopus за їх цитованістю (маленькі крапки): кожному значенню цитованості c відповідає кількість публікацій p(c). Додатково кружечками показано, який відсоток (розмір кружечка) робіт із понад 1000 співаторами міститься у кожному значенні p(c). Бачимо, що найменша частка таких колаборативних робіт міститься серед малоцитованих публікацій. Натомість більшість із високоцитованих статей мають понад 1000 співавторів.*

Як видно із Рис. 6, публікації з великими авторськими колективами, розмір яких перевищує 1000 співавторів, розподілені неоднорідно — вони переважно знаходяться у хвості розподілу. Тому природно, що нехтування цією вибіркою не просто зміщує розподіл, але змінює його форму. Саме хвіст розподілу може суттєво вплинути на середні значення, а також відіграє ключову роль у формуванні ядра індексу Гірша. Останній за означенням слабо чутливий при зміні числа малоцитованих робіт, проте сильно відчуває зміни, що стосуються кількості високоцитованих публікацій.

Оскільки ННЦ "Харківський фізико-технічний інститут" має свою специфіку, що, окрім іншого, пов'язана з великою часткою експериментальних досліджень, які виконуються на вартісному і габаритному обладнанні, то цікаво розглянути для порівняння ще приклад чисто теоретичної установи. У профілі *Інституту теоретичної фізики ім. М.М. Боголюбова НАН України (ІТФ)* у базі даних Scopus знаходимо 4004 публікації. В Табл. 4 наведені величини, що аналогічні до тих, які обговорювалися у випадку ХФТІ. Серед них, як було вже згадано, знаходимо меншу кількість (і меншу питому частку) статей із великими колективами співавторів. Так розмір найбільших стійких колаборацій обмежується на рівні колективів із близько 1000 співавторів. Загальна частка публікацій, що опубліковані колективним автором з розміром групи понад 300, складає 8.1%. Зауважимо, однак, що їх вилучення при розрахунку наукометричних характеристик зменшує середньорічну кількість цитувань установи — більше, ніж на 26%, а середню кількість цитувань на роботу — майже на 20%. Суттєво при цьому – на 19 пунктів – знижується також і груповий індекс Гірша.

*Табл. 4. Базові наукометричні показники, розраховані для Інституту теоретичної фізики ім. М.М.Боголюбова НАН України за профілем установи у базі даних Scopus станом на кінець січня 2020 р. Бралися до уваги роботи, опубліковані в період від 1991 до 2019 року включно.*

| Показник | Всі дані | Вибірка даних: статті, де кількість співавторів не перевищує *n* | | |
|---|---|---|---|---|
| | | *n*=900 | *n*=300 | *n*=50 |
| Загальна кількість публікацій | 3598 | 3481 | 3305 | 3296 |
| Частка неврахованих статей | - | 3.3% | 8.1% | 8.4% |
| Середня кількість співавторів на одну роботу | 79 | 48 | 4 | 3 |
| Середня кількість публікацій на рік | 124 | 115 | 114 | 114 |
| Середня кількість цитувань на рік | 2302 | 1751 | 1695 | 1669 |
| Середня кількість цитувань на одну роботу | 18.6 | 18.2 | 14.9 | 14.7 |
| Частка нецитованих робіт | 20% | 20% | 21% | 21% |
| Індекс Гірша | 108 | 105 | 89 | 88 |

На Рис. 7 приведені розподіли публікацій за кількістю авторів у двох установах, що розглядалися вище. Порівнюючи криві на цьому рисунку з Рис. 4, бачимо, що піки у хвості розподілів фактично повторюються, що свідчить про виняткого важливу роль цих двох академічних інститутів у великих міжнародних дослідницьких проектах за участю України.

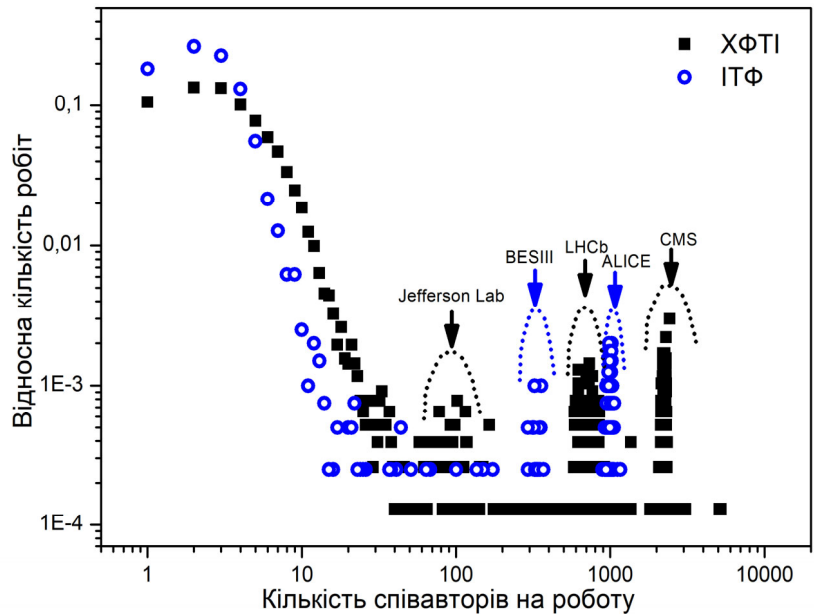

***Рис. 7.*** *Нормований частотний розподіл публікацій українських авторів за числом співавторів із результатами для двох установ: ННЦ "Харківський фізико-технічний інститут" (квадратики) та Інституту теоретичної фізики ім. М.М.Боголюбова НАН України (кружечки).*

Очевидно, що кількість установ, які причетні до масштабних міжнародних проектів і виконуються у великих колабораціях, є порівняно незначною, а сам процес формування таких колективів все ще має порівняно невелику історію. Проте, як показує уже найпростіший аналіз, сам факт їх появи вимагає суттєвого перегляду уже усталених в наукометрії підходів, які продукують показники, що часто використовуються в схемах оцінювання діяльності чи рейтингування установ (наприклад, див. [23], [24]) або ж, зокрема, окремих дослідників.

**Вплив на наукометричні профілі дослідників**
А тепер, на завершення, проаналізуємо також питання про вплив особистої участі окремих дослідників у великих стійких колабораціях на їхні індивідуальні наукометричні показники. Розглянемо, для прикладу, показники 15 дослідників з України (див. Рис. 8), які посідали найвищі позиції у переліку , що донедавна фігурував на сайті «Бібліометрика української науки» [23] (ранжування за індексом Гірша у версії Scopus).. Зауважимо, що цей сайт облікує лише тих науковців, для котрих існує відповідний профіль у Google Scholar (усі вони мають також персональний профіль у базі даних Scopus і приведене там написання прізвищ та імен використано нами у Табл. 5 та на Рис. 9) і щодо яких було відповідне звернення про їх включення до відповідної бази даних з прив'язкою до певної установи. Не коментуючи питання про застосовність індекса Гірша для рейтингування науковців без врахування інших факторів, доповнимо інформацію про дослідників із цього списку даними щодо їх участі у великих колабораціях співавторів (див. Табл. 5). Бачимо, що автори, які беруть активну участь у колабораціях є згруповані переважно у верхній частину списку. Так само легко встановити, до яких проектів переважно залучені ті чи інші дослідники, зокрема знаходимо: Б.В.Гриньов (CMS, ALICE); Г.М.Зінов'єв (ALICE); В.М.Пугач (LHCb, HERA-B); Є.С.Мартинов (ALICE); О.І.Іваницький (ALICE); І.М.Бондаренко (ORIGIN trial); В.І.Третяк (Eureca); В.В.Кобичев (Darkside, Eureca, Borexino).

| № п/п | П. І. Б. | h-index Google Scholar | Scopus | WoS | Галузь науки Рубрика Google Scholar | Установа |
|---|---|---|---|---|---|---|
| 1 | Гриньов Борис Вікторович | 133 | 81 | - | Фізика та математика Chemical & Material Sciences | Інститут сцинтиляційних матеріалів |
| 2 | Зінов'єв Геннадій Михайлович | 92 | 70 | - | Фізика та математика High Energy & Nuclear Physics | Інститут теоретичної фізики ім.М.М.Боголюбова |
| 3 | Пугач Валерій Михайлович | 105 | 65 | - | Фізика та математика High Energy & Nuclear Physics | Інститут ядерних досліджень |
| 4 | Мартинов Євген Сергійович | 76 | 59 | 58 | Фізика та математика High Energy & Nuclear Physics | Інститут теоретичної фізики ім.М.М.Боголюбова |
| 5 | Демченко Олександр Петрович | 60 | 50 | 42 | Науки про життя Nanotechnology | Інститут біохімії ім.О.В.Палладіна |
| 6 | Третяк Володимир Ілліч | 57 | 50 | 50 | Фізика та математика High Energy & Nuclear Physics | Інститут ядерних досліджень |
| 7 | Іваницький Олексій Ігорович | 50 | 48 | - | Фізика та математика High Energy & Nuclear Physics | Інститут теоретичної фізики ім.М.М.Боголюбова |
| 8 | Морозовська Ганна Миколаївна | 56 | 48 | 48 | Фізика та математика Mathematical Physics | Інститут фізики |
| 9 | Бондаренко Ігор Миколайович | 56 | 46 | 44 | Медицина Oncology | Дніпропетровська медична академія |
| 10 | Єлісєєв Євген Анатолійович | 50 | 46 | 45 | Фізика та математика Materials Engineering | Інститут проблем матеріалознавства ім.І.М.Францевича |
| 11 | Кобичев Владислав Валерієвич | 52 | 45 | 42 | Фізика та математика High Energy & Nuclear Physics | Інститут ядерних досліджень |
| 12 | Лебовка Микола Іванович | 53 | 45 | - | Хімія Dispersion Chemistry | Інститут біоколоїдної хімії ім.Ф.Д.Овчаренка |
| 13 | Ангельський Олег Вячеславович | 53 | 44 | 42 | Фізика та математика Optics & Photonics | Чернівецький національний університет імені Юрія Федьковича |
| 14 | Костюк Платон Григорович (1924-2010) | 54 | 44 | - | Науки про життя Physiology | Інститут фізіології ім.О.О.Богомольця |
| 15 | Ізотов Юрій Іванович | 56 | 43 | - | Фізика та математика Astronomy & Astrophysics | Інститут теоретичної фізики ім.М.М.Боголюбова |

***Рис. 8***. *Фрагмент зображення екрану з переліком науковців, розміщених у порядку спадання індексу Гірша за версією Scopus [23], дані станом на березень 2020 року.*

***Табл. 5***. *Наукометричні показники для дослідників, які згадані на Рис. 8. Використано дані, що стосуються їх індивідуальних профілів у базі даних Scopus станом на початок березня 2020 р.*

| | Вчений | Кількість його публікацій у Scopus | Публікаційний період | Частка публ., де є понад 50 спів-авторів | Частка публ., де є понад 500 спів-авторів | h-індекс |
|---|---|---|---|---|---|---|
| 1 | Grynyov, B | 917 | 1983-2020 | 76% | 75% | 82 |
| 2 | Zinovjev, Gennady M | 410 | 1968-2020 | 68% | 67% | 72 |
| 3 | Pugatch, V | 581 | 1991-2019 | 90% | 87% | 65 |
| 4 | Martynov, Evgenij S | 166 | 1979-2019 | 64% | 64% | 61 |
| 5 | Demchenko, Alexander P | 188 | 1969-2019 | - | - | 50 |
| 6 | Tretyak, Vl I | 239 | 1976-2019 | 13% | - | 50 |
| 7 | Ivanytskyi, Oleksii | 89 | 2012-2019 | 72% | 72% | 49 |
| 8 | Morozovska, Anna N. | 246 | 1999-2020 | - | - | 48 |
| 9 | Bondarenko, Igor N | 112 | 1984-2020 | 9% | 4% | 47 |
| 10 | Eliseev, Eugene A | 227 | 2001-2020 | - | - | 47 |
| 11 | Kobychev, Vladislav V. | 269 | 1976-2020 | 53% | - | 46 |
| 12 | Lebovka, Nikolai I | 285 | 1983-2020 | - | - | 45 |
| 13 | Angelsky, Oleg V | 248 | 1982-2019 | - | - | 44 |

| 14 | Kostyuk, Platon G | 432 | 1953-2010 | - | - | 44 |
| 15 | Izotov, Yuri I | 152 | 1989-2020 | - | - | 43 |

На Рис. 9 показана динаміка росту індекса Гірша по роках для дослідників, що згадані у Табл. 5. Для порівняння також показано, як змінювався з роками груповий індекс Гірша для трьох колективних авторів (CMS, ALICE, LHCb) на основі даних про публікації, що віднесені сервісом Scopus до України. Можна легко побачити, що темпи нарощування показника Гірша суттєво залежать від того, чи входить дослідник до колаборації великих проектів. В останньому випадку характер динаміки в значній мірі визначається саме вкладом стійких колективів співавторів із різким наростанням у ті часові проміжки, які слідують за роками, коли відповідний проект перебував в активній фазі. Так для колаборацій за проектами Великого адронного колайдера такий приріст спостерігається, починаючи приблизно із 2010 року – саме тоді, коли ця співпраця розпочиналася. Більш полога залежність є типовою для дослідників, котрі працюють в авторських колективах, що формуються у природній спосіб і відображають соціальні особливості людської співпраці загалом. Таким чином, переконуємося, що феномен колективного авторства може суттєво проявлятися і на рівні наукометричних показників окремих дослідників, а отже цей фактор слід враховувати в усіх випадках, коли їх намагаються використовувати на практиці.

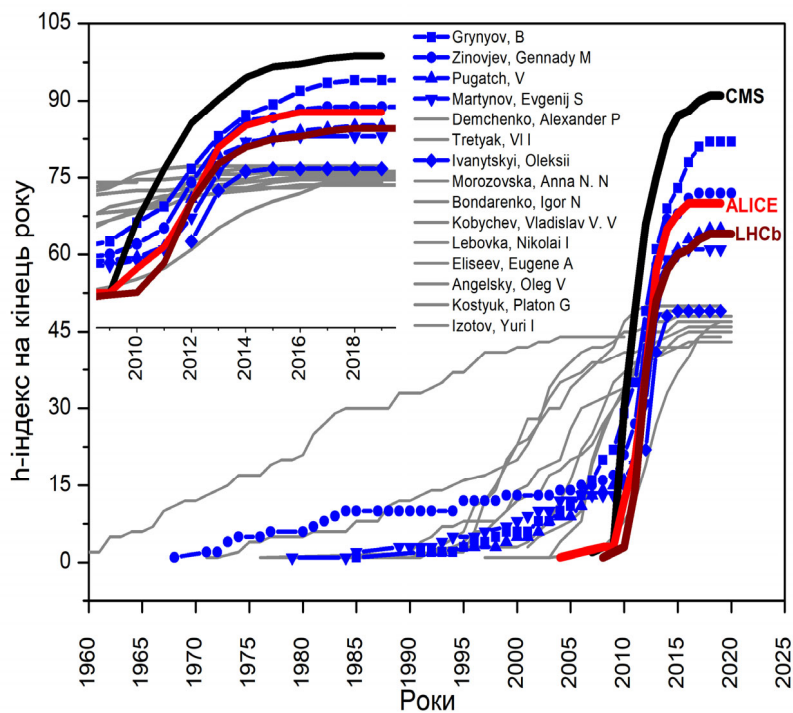

*Рис. 9. Динаміка зміни індексів Гірша для деяких дослідників з України та групових індексів Гірша для кількох колективних авторів. Вставка демонструє фрагмент рисунка для проміжку після 2010 року. Товстіші криві відповідають колективним авторам (обрано найчисельніші розміром понад 500 осіб), що працюють у проектах на Великому адронному колайдері. Синім кольором та символами показано результати для вченим, які брали участь у цих проектах.*

**Висновки**

Очевидно, що феномен надвеликих груп колективних авторів стосується не усіх ділянок науки. Тому може видаватися, що враховувати його «окремою стрічкою» – це надлишкова увага до незначного числа винятків. Проте зрозуміло, що присутність серед сукупності робіт навіть незначної кількості «колаборативних», суттєво змінює наукометричні показники як окремих установ, так і задіяних у них дослідників. За умови обмежених фінансових можливостей, коли на передній план виходять проблеми зваженого розподілу коштів, зміни на одну-дві позиції у відповідному рейтингу можуть впливати на прийняття відповідних управлінських рішень. З іншого боку, рейтингування саме по собі запускає зворотні процеси: вищі позиції у рейтингу, навіть якщо за ними стоять порівняно незначні відмінності у формальних показниках (часто в межах похибок), позитивно впливають на авторитет (і навпаки) і можуть проявлятися уже на рівні експертних оцінок. Спрацьовує так звана теорема Томаса [25] в соціології, коли суб'єктивне сприйняття дає цілком об'єктивні наслідки, і сама назва якої, свідчить у користь її змісту. Це лише один із аспектів проблеми [25], що говорить про її актуальність. Загальні тенденції росту частки публікацій, що представляють великі колаборації, стимулюють також фундаментальний інтерес до цього явища в наукометрії, а результати нашого аналізу, проведеного на прикладі українського сегменту науки, наглядно демонструють, що для його адекватного опису і коректного врахування необхідно розвивати цілком нові методи та підходи.

**Olesya Mryglod,**
Researcher of the Institute for Condensed Matter Physics of the National Academy of Sciences of Ukraine, PhD (Information Technologies)
ORCID ID: orcid.org/0000-0003-4415-7061

**Ihor Mryglod,**
Director of the Institute for Condensed Matter Physics of the National Academy of Sciences of Ukraine, Real Member of Ukr. Nat. Acad. Sci.
ORCID ID: orcid.org/0000-0002-0154-9076

## COLLECTIVE AUTHORSHIP IN UKRAINIAN SCIENCE: MARGINAL EFFECT OR NEW PHENOMENON?

One of the features of modern science is the formation of stable large collaboration of researchers working together within the projects that require the concentration of huge financial and human resources. Results of such common work are published in scientific papers by large co-authorship teams that include sometimes thousands of names. The goal of this work is to study the influence of such publications on the values of scientometric indicators calculated for individuals, research groups and science of Ukraine in general. Bibliometric data related to Ukraine, some academic Institutions and selected individual researchers were collected from Scopus database and used for our study. It is demonstrated that while the relative share of publications by collective authors is comparatively small, their presence in a general pool can lead to statistically significant effects. The obtained results clearly show that traditional quantitative approaches for research assessment should be changed in order to take into account this phenomenon.

**Keywords:** collective authorship, scientometrics, group science, Ukraine